# Enhanced Excitation and Emission from 2D Transition Metal Dichalcogenides with All–Dielectric Nanoantennas

Sergey Lepeshov[1], Alex Krasnok[2], and Andrea Alù[2,*]

[1]*ITMO University, St. Petersburg 197101, Russia*

[2]*Photonics Initiative, Advanced Science Research Center, City University of New York, New York 10031, USA*

*aalu@gc.cuny.edu*

## Abstract

The recently emerged concept of all-dielectric nanophotonics based on optical Mie resonances in high-index dielectric nanoparticles has proven a promising pathway to boost light-matter interactions at the nanoscale. In this work, we discuss the opportunities enabled by the interaction of dielectric nanoresonators with 2D transition metal dichalcogenides (2D TMDCs), leading to weak and strong coupling regimes. We perform a comprehensive analysis of bright exciton photoluminescence (PL) enhancement from various 2D TMDCs, including $WS_2$, $MoS_2$, $WSe_2$, and $MoSe_2$ via their coupling to Mie resonances of a silicon nanoparticle. For each case, we find the system parameters corresponding to maximal PL enhancement taking into account excitation rate, Purcell factor and radiation efficiency. We demonstrate numerically that all-dielectric Si nanoantennas can significantly enhance the PL intensity from 2D TMDC by a factor of hundred through precise optimization of the geometrical and material parameters. Our results may be useful for high-efficiency 2D TMDC-based optoelectronic, nanophotonic, and quantum optical devices.

## Introduction

Single-layer transition metal dichalcogenides (TMDCs) described by the general chemical formula $MX_2$ (M = Mo, W; X = S, Se) are two-dimensional (2D) semiconductors with energy bandgaps spannng the visible and near-infrared spectral region (1.0–2.1 eV), and thus raising great interest for photonics and optoelectronics applications [1–9]. They demonstrate enhancement of photoluminescence (PL) with decreasing number of layers, down to the atomic monolayer limit, which has been attributed to the transition from indirect to direct bandgap emission [9,10]. Because of weak dielectric screening, the optical properties of 2D TMDCs are governed by highly bound excitons, hydrogen-like excitation states formed by a negatively charged electron and a positively charged hole coupled through Coulomb attraction [3]. These excitations occur at the energy-degenerate K and K' endpoints of the hexagonal Brillouin zone and are characterized by large binding energies. For example, the binding energies of so-called bright ($X_A$) excitons in 2D $MoS_2$ [11], $MoSe_2$ [12], $WSe_2$ and $WS_2$ [13,14] are in the range of 0.2 to 0.8 eV, well above the thermal energy (~50 meV). In addition, due to inversion symmetry breaking and strong spin-orbital coupling in 2D TMDCs, the electronic states of the two valleys have different chirality, which leads to valley-selective circular dichroism and thus can enable valleytronic applications [4,15–17].

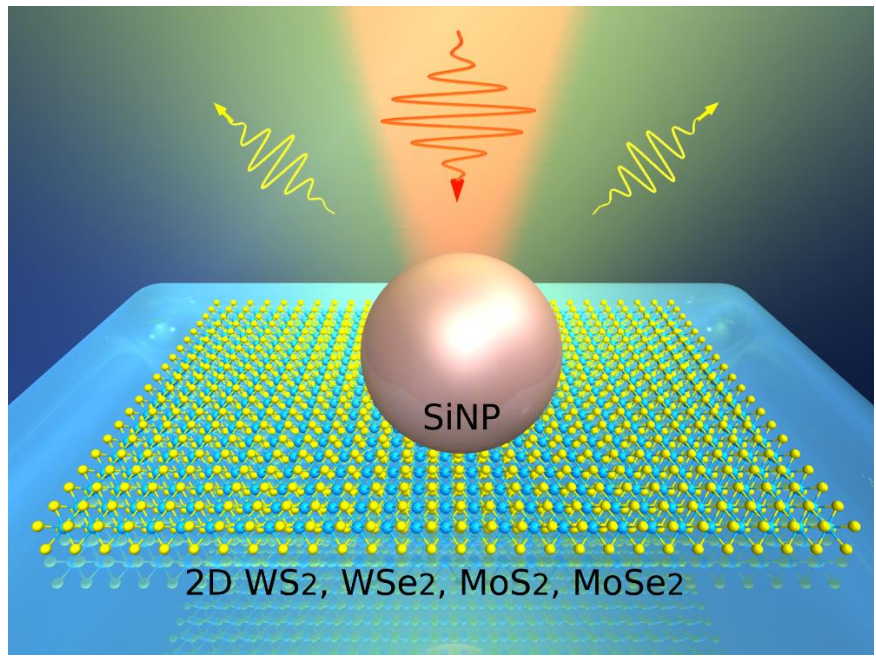

**Figure 1**. Schematic of 2D TMDC emission enhanced by coupling with an all-dielectric Si nanoparticle (SiNP) of radius R.

A variety of optical phenomena stem from the interaction of 2D TMDC with plasmonic (i.e., made of noble metals) nanoscale objects or nanoantennas, which have become a topic of extensive studies [9]. It has been demonstrated that the coupling of 2D TMDCs with plasmonic nanoantennas can lead to quantum yield enhancement of emission via Purcell effect [9,18–20], formation of exciton-polaritons in strong coupling regime [9,21–23], pronounced Fano resonances and induced optical transparency [23], and valley-selective response at room temperature [24]. However, all these effects suffer from high dissipative losses, inherent for metals at optical frequencies [25–27], which cause Joule heating of the structure. Since the response of excitonic systems is dramatically modified by temperature changes, heat generation is detrimental for 2D TMDCs. These losses are caused by the large free-carrier concentration $\approx 10^{23}$ cm$^{-3}$ in plasmonic materials. Moreover, hot-electron injection from plasmonic nanostructures to TMDCs may lead to phase change and unwanted doping of TMDCs, which are not desirable in many applications [28].

All-dielectric nanophotonics, based on Mie resonances in high-index and low-loss dielectric nanoparticles (cSi, GaP, Ge) [29,30], provides an interesting alternative to induce efficient and low-loss light-matter interactions at the nanoscale [31]. Although these nanoantennas are currently of great interest for spectroscopy, bioimaging and flat optics, only a small number of papers have been exploring their interaction with 2D TMDCs [32,33], and the optimal values of the key parameters governing this interaction remain unknown. In this work, we perform a comprehensive analysis of bright exciton photoluminescence (PL) enhancement from different 2D TMDCs via coupling to Mie resonances of silicon nanoparticle (SiNP). We restrict ourselves to the visible spectral range, being the most interesting from the practical viewpoint, and hence consider 2D $WS_2$, $MoS_2$, $WSe_2$, and $MoSe_2$, whose PL spectra span the range 600–800 nm. For each material, we define the system parameters (SiNP radius, excitation and emission wavelengths) corresponding to maximum PL enhancement, taking into account excitation rate, Purcell factor and radiation efficiency. We also analyze the power patterns of PL emission that allows an estimate of the emission towards different angles.

## Results and Discussion

The system under study consists of 2D TMDC ($WS_2$, $MoS_2$, $WSe_2$, and $MoSe_2$) on a $SiO_2$ substrate, corresponding to the typical geometry in experiments. We assume a Si nanoparticle (SiNP) of radius R placed on 2D TMDC (Figure 1). In particular, we consider c-Si for the particle material, because of its large refractive index in the visible range and moderate losses [30]. SiNPs of spherical shape can be fabricated chemically [33] or with fs-laser ablation [34]. In [33],[32], it has been shown that spherical SiNPs and Si nanowires can be placed on 2D TMDCs ($WS_2$ and $MoS_2$, respectively) without damaging their crystalline structure, where an additional thin layer of h-BN can be used for protection or isolation of 2D TNDCs [35].

The optical response of 2D TMDC is governed by different exciton resonances. The crystal lattice inversion symmetry breaking combined with strong spin-orbit coupling results in a large valence band splitting at the K (K') points in the first Brillouin zone. This gives rise to two different valley excitons, $X_A$ (bright) and $X_B$ (dark) ones, which are associated with optical transitions from the upper and lower valence bands to the bottom of the conduction band, respectively. This causes the appearance of two resonances in the absorption and PL spectra of 2D TMDCs. Although, $X_B$ excitons play a crucial role in the exciton relaxation dynamics, their contribution to the absorption and PL emission is very weak, especially at a room temperature. For this work, the room temperature absorption and PL spectra of WS$_2$, MoS$_2$, WSe$_2$, and MoSe$_2$ have been extracted from Refs.[18,36–38], Figures 2b and 3b, respectively.

In order to estimate the total PL intensity enhancement, we have to first consider the effect of excitation rate [31]. In the case of coherent excitons, the PL intensity $I_{PL}$ is proportional to the squared total dipole $D$ moment of all excitons involved in the coupling with SiNP, i.e., $I_{PL} \sim D^2 \approx (d \cdot N)^2$. Here, $N$ is the number of excited excitons and $d$ is the dipole moment of one exciton. However, for spontaneous emission, the assumption that the total PL intensity is the sum of the PL intensities from all excitons without interference, $I_{PL} \sim d^2 \cdot N$, is more natural. Thus, for further analysis we assume the excitons to be incoherent. Next, the number of excitons $N$ is proportional to the tangential electric field intensity ($|\mathbf{E}_t|^2$) on the TMDC surface for a given internal quantum efficiency [31]. Thus, enhanced local electric field intensity induces an increase in the number of excitons involved in the light-matter interaction process. Figure 2a shows the enhancement of the local tangential electric field intensity in the gap between SiNP and silica substrate as a function of R and the excitation wavelength. We see that the absorption maximum depends on the SiNP radius and it can be tuned to any resonant peak in the 2D TMDCs absorptivity spectra (Figure 2b). For instance, 2D WS$_2$ with an absorption maximum at 615 nm reaches the maximum in excitation rate for R = 76 nm (MD), 110 nm (MQ), 138 nm (EQ), 142 nm (MO), 173 nm (EO). We also note that the largest $|\mathbf{E}_t|^2$ is expected at the MQ resonance, as it can be expected considering the vector E-field distribution profiles for selected modes presented in Figure 2c. In fact, in contrast to electric Mie modes, magnetic modes have a tangential component of electric field to the particle surface that coupled efficiently to the bright excitons of 2D TMDCs. For example, the electric dipole (ED) mode, polarized along the radial direction, does not contribute to $|\mathbf{E}_t|^2$ (as well as the Purcell factor spectra, Figure 3). Next, although the MD mode has a tangential E-field component, its hot-spots are located at the particle poles with only moderate enhancement in the gap. This is different for the MQ mode. The MO and EO modes are localized more within the SiNP, which leads to higher losses and reduces $|\mathbf{E}_t|^2$.

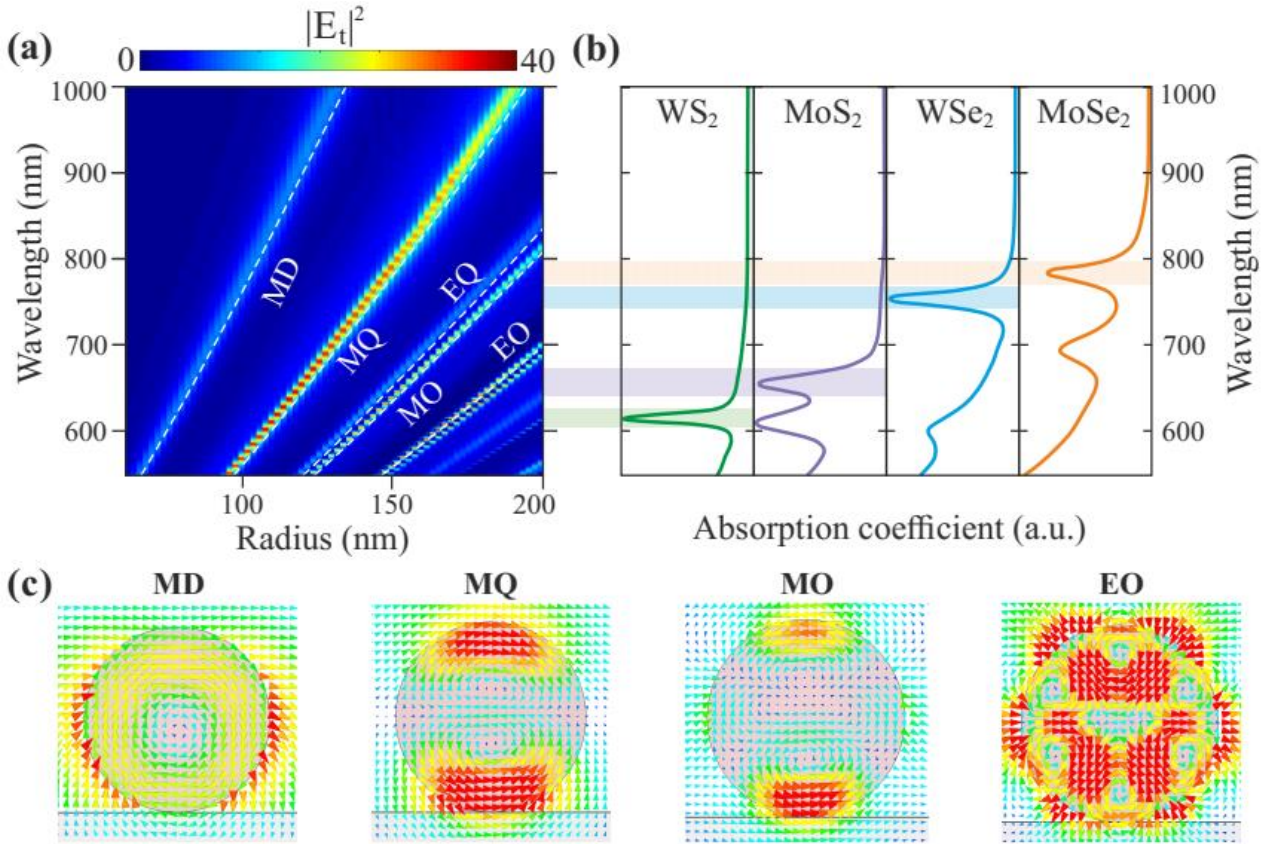

**Figure 2**. (a) Enhancement of the tangential electric field intensity $|\mathbf{E}_t|^2$ in the gap between SiNP and silica as a function of nanoparticle radius and excitation wavelength. (b) Absorption spectra of different 2D TMDC (at room temperature), arranged in order of maximum absorption wavelength (adapted from Refs. [18,36–38]). (c) Electric field distribution profiles of different modes excited in the SiNP on SiO$_2$ calculated at the wavelength of 615 nm and different radii R = 76 nm (MD), 110 nm (MQ), 142 nm (MO), and 173 nm (EO). Here MD, MQ, EQ, MO, and EO stand for magnetic dipole, magnetic quadrupole, electric quadrupole, magnetic octupole, and electric octupole, respectively.

Excitons in 2D TMDC can interact with SiNP in two possible coupling regimes, i.e., weak and strong coupling. In Ref. [33] it was shown that the system under consideration (Figure 1) has an exciton dipole moment not large enough to induce strong coupling for a given loss rate in SiNP and air as the surrounding medium. Thus, in this regime the system response is accurately described by the Purcell factor [31]. The Purcell factor $F$ quantitatively expresses the modification of the spontaneous emission rate $\gamma$ of the exciton induced by its interaction with the environment (nanoantenna), $F = \gamma / \gamma^0$. Here, $\gamma^0$ stands for the spontaneous emission rate in free-space. In steady-state under CW excitation, this quantity is simply equal to the enhancement of total emitted power $P_t$, taking into account radiation to farfield $P_{rad}$ and local absorption $P_{abs}$ in the SiNP, $F = (P_{rad} + P_{abs})/P^0$, where $P^0$ stands for the emitted power in absence of the nanoantenna. Figure 3a shows the calculated total Purcell factor of a tangential dipole emitter placed in the gap between SiNP and silica substrate as a function of R and emission wavelength.

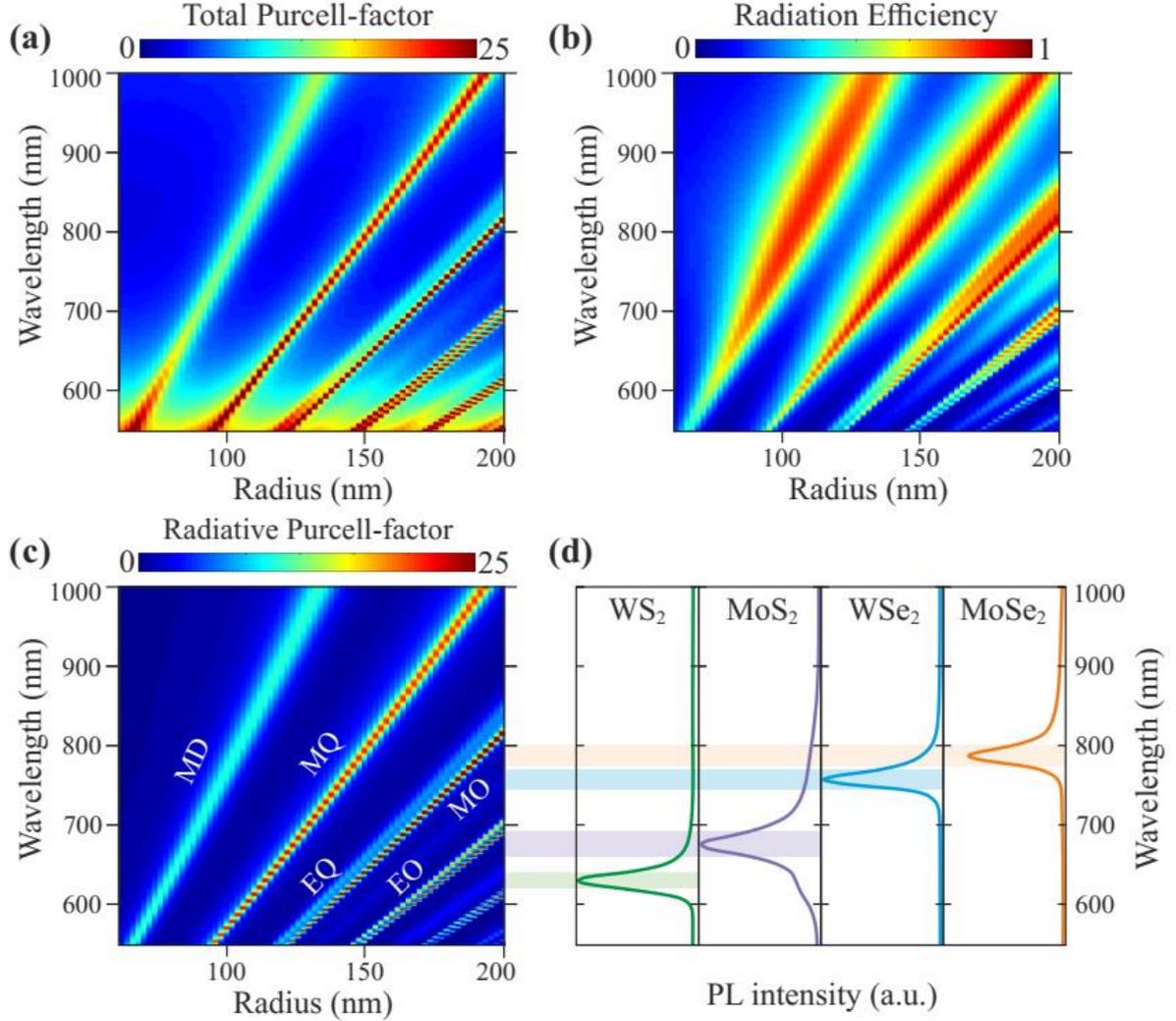

**Figure 3**. Total Purcell factor (a), radiation efficiency (b), and radiation enhancement (c) of a tangential dipole emitter (bright exciton) placed in the gap between SiNP and silica substrate as a function of the SiNP radius and emission wavelength. (d) Photoluminescence spectra of different 2D TMDC (taken at room temperature), arranged in order of PL maximum wavelength increasing (adopted from Refs.[18,36–38]).

In some circumstances a larger Purcell factor is achieved at the cost of high dissipative losses. Hence, in order to account for the effect of loss we calculate the radiation efficiency ($\eta = P_{rad}/P_t$), Figure 3b. The resulting radiation enhancement (radiative Purcell factor) is presented in Figure 3c. This result shows that the radiative Purcell factor maximum can be adjusted with any 2D TMDC PL maxima (Figure 3d) by changing the radius. For instance, 2D WS$_2$ possesses a PL maximum at 633 nm and it reaches a maximum of radiation enhancement for R = 76 nm (MD), 110 nm (MQ), 142 nm (EQ), 147 nm (MO) and 173 nm (EO).

Now we look for the maximum PL enhancement of X$_A$ excitons in 2D TMDCs, taking into account the excitation rate, Purcell factor and radiation efficiency considered above. As an example, let us consider the case of WS$_2$. If we fix the excitation and PL emission wavelength to 615 nm and 633 nm, respectively, the calculated tangential electric field intensity and radiation enhancement for different SiNP radius are shown in Figure 4a (bottom). The product of these two curves provides the total enhancement achievable for a specific R, presented in Figure 4a (top). The total enhancement has four pronounced maxima in the considered range, around the MD, MQ, EQ/MO and EO resonances with values of 30, 83, 78, and 23, respectively. It is noted that, in the case of R=142 nm, the excitation of WS$_2$ occurs at the MO resonance, while the radiation proceeds into the EQ mode. Similarly, in the case of R=173 nm, the excitation and radiation channels belong to different modes. The corresponding results for all four TMDCs are summarized in Figures 4b and 5. The largest total enhancement is achieved for materials with smallest Stokes shift (WSe$_2$ and MoSe$_2$) with values of 200 and 177, respectively (see Figures 5b and c). Note that the results for MoS2 in the vicinity of the MQ resonance qualitatively agree with the results in [32] for a Si nanowire possessing a similar mode composition.

The nanoantenna can also facilitate outcoupling of the PL emission from the system towards a desired direction. To address this issue we calculate the directivity power pattern $D(\varphi,\theta)$, defined as $D(\varphi,\theta) = 4\pi p(\varphi,\theta)/P_{rad}$, where $p(\varphi,\theta)$ is the power flow towards the direction $(\varphi,\theta)$. The results are presented in Figure 4c. We can conclude that, for a given emission wavelength, the SiNP radius can be optimized to achieve directional emission towards the desired direction.

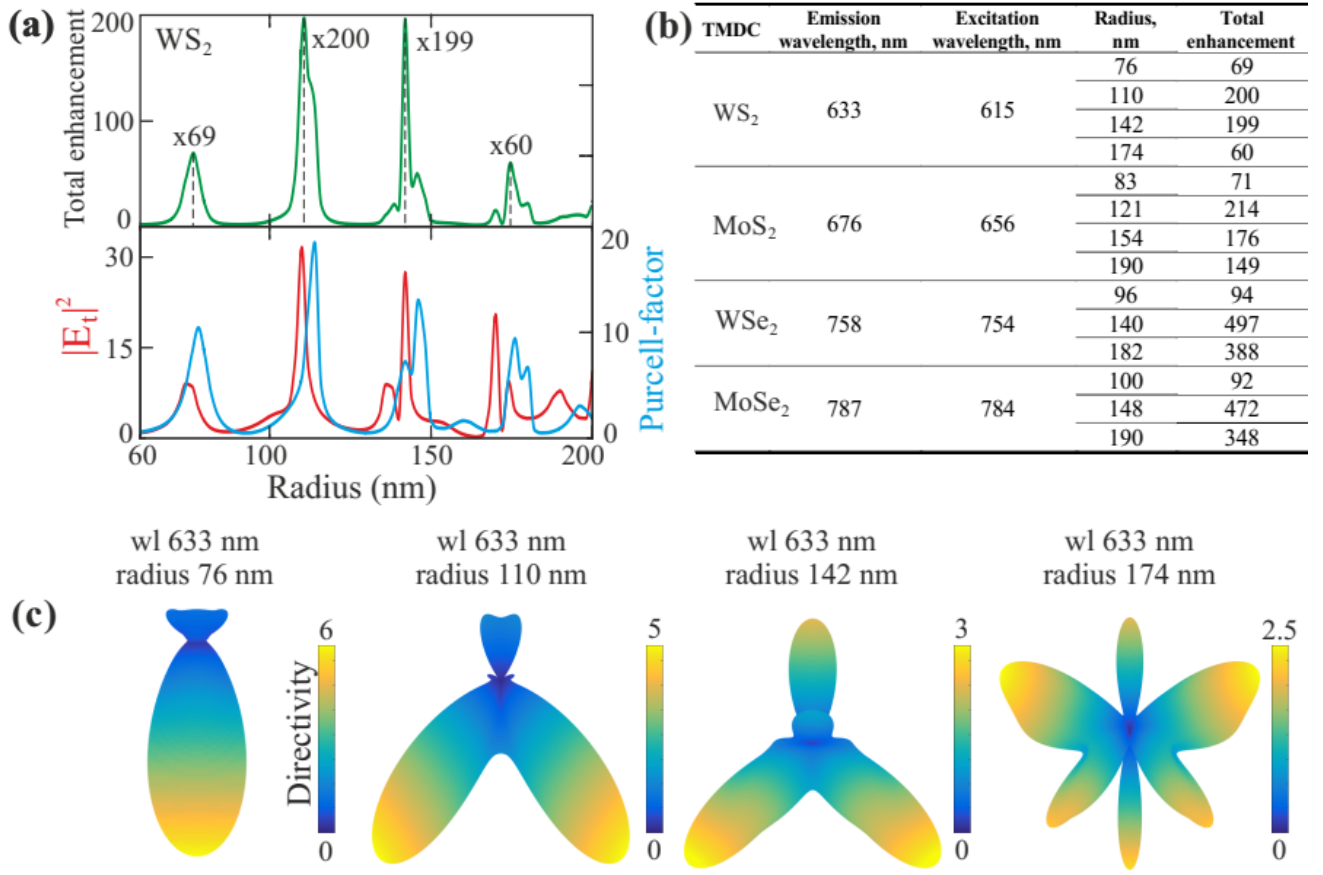

**Figure 4**. (a) Tangential electric field intensity enhancement (red curve) in the gap for WS$_2$ (excitation wavelength is 615 nm), the radiation enhancement (blue curve) of a dipole source (emission wavelength is 633 nm), and the total emission enhancement (green curve) as a function of R. (b) Comparative table of total PL enhancements for different TMDCs. (c) Power patterns calculated at the WS$_2$ emission wavelength for radii corresponding to the total enhancement peaks.

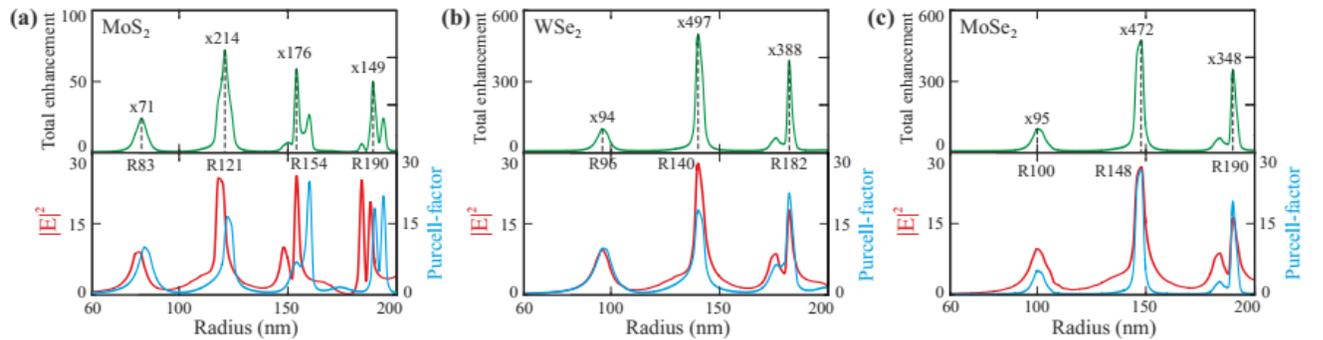

**Figure 5**. Electric field intensity enhancement (red curve) in the gap at the optimal excitation wavelength, as well as the radiation enhancement (blue curve) of a dipole source (at the PL maximum in the each specific case), and the total emission enhancement (|green curve) as a function of R for MoS$_2$ (a), WSe$_2$ (b) and MoSe$_2$ (c).

## Conclusions

In this paper, we have performed a comprehensive analysis of bright exciton photoluminescence enhancement from various 2D TMDCs, including $WS_2$, $MoS_2$, $WSe_2$, and $MoSe_2$ via coupling to Mie resonances of a silicon nanoparticle. We have demonstrated that all-dielectric Si nanoantennas can significantly enhance the intensity of the photoluminescence signal coming from 2D TMDC by two orders of magnitude, by thoughtful optimization of the system key parameters, SiNP radius, excitation and PL emission wavelengths. For each specific case, we have found the geometric parameters corresponding to maximal PL enhancement, taking into account the excitation rate, Purcell factor and radiation efficiency. We have also demonstrated that, in contrast to electric Mie modes, the magnetic modes have a tangential component of electric field to the particle surface that can couple efficiently to the bright excitons of 2D TMDCs. Finally, we have found that the MQ mode demonstrates the best performance for Si nanoantenna in the visible range, because it optimizes a tradeoff between tangential E-field component enhancement in the gap and mode confinement within the particle. Our results provide interesting insights into nanophotonic engineering of exciton emission, with opportunities for imaging, sensing and valleytronic applications.

## Acknowledgements

We acknowledge support from the Air Force Office of Scientific Research and the Welch Foundation with grant No. F-1802.